\documentclass{article}
\usepackage{amsfonts,amssymb,amsmath,graphicx,hyperref}
\input{tcilatex}

\begin{document}

\title{New Mechanics of Traumatic Brain Injury}
\author{Vladimir G. Ivancevic\\
Land Operations Division\\ Defence Science \& Technology
Organisation\\ Australia}\date{}\maketitle

\begin{abstract}
The prediction and prevention of traumatic brain injury is a very
important aspect of preventive medical science. This paper
proposes a new \emph{coupled loading--rate hypothesis} for the
traumatic brain injury (TBI), which states that the main cause of
the TBI is an \textsl{external Euclidean jolt}, or $SE(3)-$jolt,
an impulsive loading that strikes the head in several coupled
degrees-of-freedom simultaneously. To show this, based on the previously defined
\emph{covariant force law}, we formulate the coupled Newton--Euler
dynamics of brain's micro-motions within the cerebrospinal fluid
and derive from it the coupled $SE(3)-$jolt dynamics. The
$SE(3)-$jolt is a cause of the TBI in two forms of brain's rapid
discontinuous deformations: translational dislocations and
rotational disclinations. Brain's \emph{dislocations and
disclinations}, caused by the $SE(3)-$jolt, are described using
the Cosserat multipolar viscoelastic continuum brain model.
\bigbreak

\end{abstract}

\section{Introduction}

Traumatic brain injury (TBI) continues to be a major health
problem, with over 500,000 cases per year with a societal cost of
approximately \$85 billion in the US. Motor vehicle accidents are
the leading cause of such injuries. In many cases of TBI
widespread disruption of the axons occurs through a process known
as diffuse axonal injury (DAI) or traumatic axonal injury (TAI)
\cite{Singh06}. TBI occurs when physical trauma
causes brain damage, which can result from a closed head injury\footnote{%
A closed injury occurs when the head suddenly and violently hits an object
but the object does not break through the skull.} or a penetrating head
injury.\footnote{%
A penetrating injury occurs when an object pierces the skull and
enters brain tissue.} In both cases, TBI is caused by rapid
deformation of the brain, resulting in a cascade of pathological
events and ultimately neuro-degeneration. Understanding how the
biomechanics of brain deformation leads to tissue damage remains a
considerable challenge \cite{Morrison06}.

Parts of the brain that can be damaged include the cerebral
hemispheres, cerebellum, and brain stem. TBI can cause a host of
physical, cognitive, emotional, and social effects \cite{NIH02,Rapp}.
Half of all TBIs are due to transportation accidents involving
automobiles, motorcycles, bicycles, and pedestrians. These
accidents are the major cause of TBI in people under age 75. For
those aged 75 and older, falls cause the majority of TBIs.
Approximately 20 \% of TBIs are due to violence, such as firearm
assaults and child abuse, and about 3\% are due to sports
injuries. Fully half of TBI incidents involve alcohol use
\cite{NIH02}. TBI is a frequent cause of major long-term
disability in individuals surviving head injuries sustained in war
zones. This is becoming an issue of growing concern in modern
warfare in which rapid deployment of acute interventions are
effective in saving the lives of combatants with significant head
injuries. Traumatic brain injury has been identified as the
`signature injury' among wounded soldiers of the current military
engagement in Iraq \cite{Mason07,Hoge08}. Rapid deformation of
brain matter caused by skull acceleration is most likely the cause
of concussion, as well as more severe TBI. The inability to
measure deformation directly has led to disagreement and confusion
about the biomechanics of concussion and TBI \cite{Bayly05}.

TBI can be mild, moderate, or severe, depending on the extent of
the damage to the brain. Outcome can be anything from complete
recovery to permanent disability or death (see \cite{Chen}). Some symptoms are
evident immediately, while others do not surface until several
days or weeks after the injury \cite{NIH02}. With mild TBI, the
patient may remain conscious or may lose consciousness for a few
seconds or minutes; the person may also feel dazed or not like
him- or herself for several days or weeks after the initial
injury; other symptoms include: headache, mental confusion,
lightheadedness, dizziness, double vision, blurred vision (or
tired eyes), ringing in the ears, bad taste in the mouth, fatigue
or lethargy, a change in sleep patterns, behavioral or mood
changes, trouble with memory/concentration/calculation. With
moderate or severe TBI, the patient may show these same symptoms,
but may also have: loss of consciousness, personality change, a
severe/persistent/worsening headache, repeated vomiting/nausea,
seizures, inability to awaken, dilation (widening) of one or both
pupils, slurred speech, weakness/numbness in the extremities, loss
of coordination, increased confusion, restlessness/agitation;
vomiting and neurological deficit together are important
indicators of prognosis and their presence may warrant early CT
scanning and neurosurgical intervention.

In particular, standard medical statistics suggest that the loss
of consciousness in boxing knock-outs and road--vehicle crashes is
caused by \textit{rotation of the brain--stem} \cite{Misra1} as a
dynamic response \cite{12,14a} of a head--neck system to an
impulsive load \cite{Misra2}. It is generally associated to the
following three syndromes:\textit{\ Locked-In, Semi-Coma,} and
\textit{Akinetic Mute}, all three characterized by the total loss
of gesture, speech and movement. The cognitive abilities \cite
{20,LifeSpace} can still be intact, but the patient cannot express
himself by either speech or gesture. Recall that the brain stem,
including Midbrain, Pons and Medulla Oblongata, is located at the
base of the brain. It is the connection between the cortex and the
spinal cord, containing motor neural pathways for voluntary
movement from the upper part of the brain. The brain stem also
controls such automatic functions as breathing, heart rate, blood
pressure, swallowing, sleep patterns and body temperature. Weaker
injuries include another three symptoms: \textit{abnormal
respiration} (hyperventilation and abnormal breathing patterns:
ataxic, clustered, hiccups); \textit{pupils}: dilated, fixed; and
\textit{movement} (if any): abnormal extensor.

\begin{figure}[tbh]
\centering \includegraphics[width=14cm]{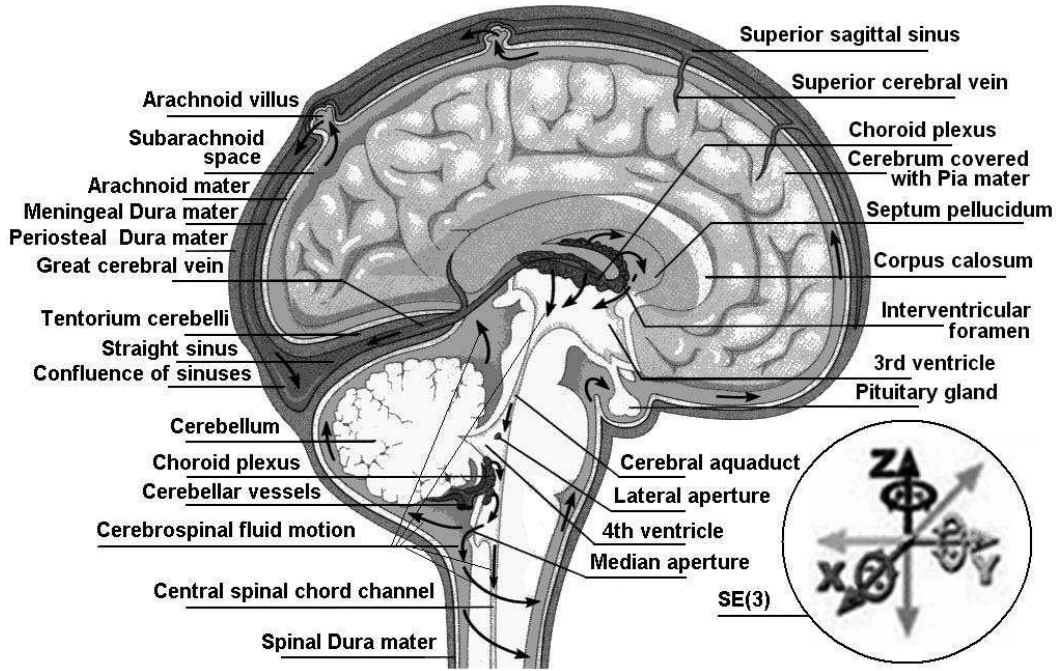} \caption{Human
brain and its SE(3)--group of microscopic three-dimensional (3D)
motions within the cerebrospinal fluid inside the cranial cavity.}
\label{IvBrain999}
\end{figure}

The natural cushion that protects the brain from trauma is the
\emph{cerebrospinal fluid} (CSF). It resides within cranial and
spinal cavities and moves in a pulsatile fashion to and from the
cranial cavity (see Figure \ref{IvBrain999}). This motion can be
measured by functional magnetic resonance imaging (fMRI, see \cite{Sokoloff} for a review) and may be of
clinical importance in the diagnosis of several brain and spinal
cord disorders such as hydrocephalus, Chiari malformation, and
syringomyelia. It was found in \cite{Maier94} that brain and CSF
of healthy volunteers exhibited periodic motion in the frequency
range of normal heart rate. Both brain hemispheres showed periodic
squeezing of the ventricles, with peak velocities up to 1 mm/sec
followed by a slower recoil. Superimposed on the regular
displacement of the brain stem was a slow, respiratory-related
periodic shift of the neutral position. During the Valsalva
maneuver, the brain stem showed initial caudal and subsequent
cranial displacement of 2-3 mm. Coughing produced a short swing of
CSF in the cephalic direction. The pressure gradient waveform of a
linearized Navier-Stokes model of the pulsatile CSF flow was found
in \cite{Loth01} to be almost exclusively dependent on the flow
waveform and cross-sectional area.

The state of head injury biomechanics:: past, present, and future
was presented in \cite{Goldsmith01}, dealing with components and
geometry of the human head, classification of head injuries,
tolerance considerations, head motion and load characterization,
experimental dynamic loading of human living and cadaver heads,
dynamic loading of surrogate heads, and head injury mechanics. The
subsequent paper \cite{Goldsmith05} described physical head injury
experimentation involving animals (primarily primates), human
cadavers, volunteers, and inanimate physical models.

Besides motor accidents, concussion (or mild TBI), occurs in many
activities, mostly as a result of the head being accelerated. For
example, although a popular endeavor, boxing has fallen under
increased scrutiny because of its association with TBI. The injury
rate in professional boxing matches is high, particularly among
male boxers. Superficial facial lacerations are the most common
injury reported. Male boxers have a higher rate of knockout and
technical knockouts than female boxers \cite{Bledsoe05}. Although
the epidemiology and mechanics of concussion in sports have been
investigated for many years, the biomechanical factors that
contribute to mild TBI remain unclear because of the difficulties
in measuring impact events in the field. The objective of
\cite{Beckwith07} was to validate an instrumented boxing headgear
(IBH) that can be used to measure impact severity and location
during play. Based on this study, the IBH is a valid system for
measuring head acceleration and impact location that can be
integrated into training and competition. Similarly, a
comprehensive study has been conducted by \cite{Newman05} to
understand better the mechanics of the impacts associated with
concussion in American football, involving a sequence of
techniques to analyze and reconstruct many different head impact
scenarios.

Impact biomechanics from boxing punches causing translational and
rotational head acceleration was experimentally studied in
\cite{Viano05}. Olympic boxers threw four different punches at an
instrumented Hybrid III dummy and responses were compared with
laboratory-reconstructed NFL concussions. Instrumentation included
translational and rotational head acceleration and neck loads in
the dummy. Biaxial acceleration was measured in the boxer's hand
to determine punch force. Hybrid III dummy head responses and FE
brain modelling were compared to similarly determined responses
from reconstructed concussions in professional NFL football
players. The hook produced the highest change in hand velocity
(11.0 +/- 3.4 m/s) and greatest punch force (4405 +/- 2318 N) with
average neck load of 855 +/- 537 N. It caused head translational
and rotational accelerations of 71.2 +/- 32.2 g and 9306 +/- 4485
$r/s^2$. These levels are consistent with those causing concussion in
NFL impacts. However, the head injury criterion (HIC) for boxing
punches was lower than for NFL concussions because of shorter
duration acceleration. Boxers deliver punches with proportionately
more rotational than translational acceleration than in football
concussion. Boxing punches have a 65 mm effective radius from the
head cg, which is almost double the 34 mm in football. A smaller
radius in football prevents the helmets from sliding off each
other in a tackle. Similarly, impacts causing concussion in
professional football were simulated in laboratory tests to
determine collision mechanics \cite{Viano07}. This study focused
on the biomechanics of concussion in the struck player, addressing
head responses causing concussion in the NFL players.

Head injury mechanisms are difficult to study experimentally due
to the variety of impact conditions involved, as well as ethical
issues, such as the use of human cadavers and animals\\
\cite{Krabbel95}. A number of finite element (FE) models and
analysis studies have been conducted in order to understand the
mechanism of TBI. An FE analysis was carried out in \cite{Chu94}
to study the mechanism of cerebral contusion. Clinical findings
indicate that most cerebral contusions in the absence of skull
fracture occur at the frontal and temporal lobes. To explain these
observations, cavitation and shear strain theories have long been
advocated. Plane strain finite element models of a para-sagittal
section of the human head were developed in the present study. The
model was first validated against a set of experimental results
from the literature. Frontal and occipital impacts were then
simulated, and pressure and shear stress distributions in the
brain were compared. While comparable negative pressures always
developed in the contrecoup regions, shear stress distributions
remained nearly identical regardless of the impact direction,
consistent with the clinically observed pattern for contusion.
Therefore, shear strain theory appeared to account better for the
clinical findings in cerebral contusion.

A 3D FE model based on the anatomical features of the adult human
cranium was developed in \cite{Krabbel95}. The complex cranial
geometry was measured from a series of 2D computer tomography
images. The CT scans were transformed with a self-developed
preprocessor into a finite element mesh. A review of the existing
FE models was presented in \cite{Voo96} for the biomechanics of
human head injury. More recent models incorporated anatomic
details with higher precision. The cervical vertebral column and
spinal cord were included. Model results had been more qualitative
than quantitative owing to the lack of adequate experimental
validation. Advances included transient stress distribution in the
brain tissue, frequency responses, effects of boundary conditions,
pressure release mechanism of the foramen magnum and the spinal
cord, verification of rotation and cavitation theories of brain
injury, and protective effects of helmets. These theoretical
results provided a basic understanding of the internal
biomechanical responses of the head under various dynamic loading
conditions. The mechanism of brain contusion has been investigated
in in \cite{Huang00} using a series of 3D FE analyses. A head
injury model was used to simulate forward and backward rotation
around the upper cervical vertebra. Intracranial pressure and
shear stress responses were calculated and compared. The results
obtained with this model support the predictions of cavitation
theory that a pressure gradient develops in the brain during
indirect impact. Contrecoup pressure-time histories in the
para-sagittal plane demonstrated that an indirect impact induced a
smaller intracranial pressure (-53.7 kPa for backward rotation,
and -65.5 kPa for forward rotation) than that caused by a direct
impact. Comparison of brain responses between frontal and lateral
impacts was performed in \cite{Zhang01} by FE modelling. Identical
impact and boundary conditions were used for both the frontal and
lateral impact simulations. Intracranial pressure and localized
shear stress distributions predicted from these impacts were
analyzed. The model predicted higher positive pressures
accompanied by a relatively large localized skull deformation at
the impact site from a lateral impact when compared to a frontal
impact. Lateral impact also induced higher localized shear stress
in the core regions of the brain.

A nonlinear viscoelastic FE model for brain tissue was developed
in \cite{Brands04}. To obtain sufficient numerical accuracy for
modelling the nearly incompressible brain tissue, deviatoric and
volumetric stress contributions were separated. Deviatoric stress
was modelled in a nonlinear viscoelastic differential form. An
attempt was made in \cite{Zhang04} to delineate actual injury
causation and establish a meaningful injury criterion through the
use of the actual field accident data. Twenty-four head-to-head
field collisions that occurred in professional football games were
duplicated using a validated FE human head model. The injury
predictors and injury levels were analyzed based on resulting
brain tissue responses and were correlated with the site and
occurrence of mild TBI. Predictions indicated that the shear
stress around the brainstem region could be an injury predictor
for concussion. The controlled cortical impact model has been used
extensively to study focal traumatic brain injury. Although the
impact variables can be well defined, little is known about the
biomechanical trauma as delivered to different brain regions. The
FE analysis based on high resolution T2-weighted MRI images of rat
brain was used in \cite{Pena05} to simulate displacement, mean
stress, and shear stress of brain during impact. Young's Modulus
E, to describe tissue elasticity, was assigned to each FE in three
scenarios: in a constant fashion (E = 50 kPa), or according to the
MRI intensity in a linear (E = [10, 100] kPa) and inverse-linear
fashion (E = [100, 10] kPa). Simulated tissue displacement did not
vary between the 3 scenarios, however mean stress and shear stress
were largely different. The linear scenario showed the most likely
distribution of stresses.

A detailed FE model of the rat brain was developed in \cite{Mao06}
for the prediction of intracranial responses due to different
impact scenarios. The FE model was used to predict biomechanical
responses within the brain due to controlled cortical impacts
(CCI). A total of six different series of CCI studies, four with
unilateral craniotomy and two with bilateral craniotomy, were
simulated and the results were systematically analyzed, including
strain, strain rate and pressure within the rat brain. Simulation
results indicated that intracranial strains best correlated with
experimentally obtained injuries. An automating meshing method for
patient-specific FE model was developed in \cite{Guan06}. 3D
geometries of two six-month-old infant heads were reconstructed
from the CT data. FE meshes including cranial bone of skull,
brain, and suture were generated. Both static and dynamic analyzes
were performed to verify the models. The study for blunt impact of
infant head was performed by using these patient-specific models.

A 3D FE analysis of human head was performed in \cite{Zong06}, to
assess injury likelihood of the head subjected to impact loading.
The structural intensity (SI) methodology\footnote{SI is a vector
quantity indicating the direction and magnitude of power flow
inside a dynamically loaded structure.} was introduced in
accordance with the prevailing practice in experimental
biomechanics. The SI field inside the head model was computed for
three frontal, rear and side impacts. The results for the three
cases revealed that there existed power flow paths. The skull was,
in general, a good energy flow channel. This study also revealed
the high possibility of spinal cord injury due to wave motion
inside the head. Recently, a 3D FE analysis was performed in
\cite{Takahashi07} in respect to the frequency analysis of the
pressure changes related to TBI. From the results of computer
simulations and impact experiments, the authors found similar
spectrums in some frequency bands, which indicated the occurrence
of the brain injury. A vigorous shaking and an inflicted impact
were compared in \cite{Roth07}, defined as the terminal portion of
a vigorous shaking, using a FE model of a 6-month-old child head.
Whereas the calculated values in terms of shearing stress and
brain pressure remain different and corroborate the previous
studies based on angular and linear velocity and acceleration, the
calculated relative brain and skull motions that can be considered
at the origin of a subdural haematoma show similar results for the
two simulated events. A 2D FE model was developed in \cite{Li07}
with objective to determine localized brain's strains in lateral
impact using finite element modelling and evaluate the role of the
falx. Motions and strains from the stress analysis matched well
with experimental results from literature. A parametric study was
conducted by introducing flexible falx in the finite element
model. For the model with the rigid falx, high strains were
concentrated in the corpus callosum, whereas for the model with
the flexible falx, high strains extended into the cerebral vertex.

On the other hand, it has been a common perception that
\emph{rapid head rotation} is a major cause of brain damage in
automobile crashes and falls. A model for \emph{rotational
acceleration} about the center of mass of the rabbit head was
presented in \cite{Gutierrez01}, which allowed the study of brain
injury without translational acceleration of the head. In the
companion paper \cite{Runnerstam01} it was shown that rotational
head acceleration caused extensive subarachnoid hemorrhage, focal
tissue bleeding, reactive astrocytosis, and axonal damage. The
initial response of the brain after rotational head injury
involved brain edema after 24 h and an excitotoxic neuronal
micro-environment in the first hour, which leaded to extensive
delayed neuronal cell death by apoptosis necrosis in the cerebral
cortex, hippocampus and cerebellum. Similarly, the study by
\cite{Zhang06} used the SIMon human FE head model and delineated
the contributions of these accelerations using post mortem human
subject (PMHS) lateral head impact experimental data. Results
indicated that rotational acceleration contributed more than 90\%
of total strain, and translational acceleration produced minimal
strain.

The fidelity of cell culture TBI--simulations that yield tolerance
and mechanistic information relies on both the cellular models and
mechanical insult parameters. An electro-mechanical cell shearing
device was designed by \cite{LaPlaca05} in order to produce a
controlled high strain rate injury (up to 0.50 strain, 30 s(-1)
strain rate) that deforms 3D neural cultures (neurons or
astrocytes in an extracellular matrix scaffold). Theoretical
analysis revealed that these parameters generated a heterogeneous
3D strain field throughout the cultures that was dependent on
initial cell orientation within the matrix, resulting in various
combinations of normal and shear strain.

Rigid-body modelling (RBM) was used in \cite{Wolfson05} to
investigate the effect of neck stiffness on head motion and
head-torso impacts as a possible mechanism of injury. Realistic
shaking data obtained from an anthropometric test dummy (ATD) was
used to simulate shaking. In each study injury levels for
concussion were exceeded, though impact-type characteristics were
required to do so in the neck stiffness study. Levels for the type
of injury associated with the syndrome were not exceeded.

The nonlinear mechanical behavior of porcine brain tissue in large
shear deformations was determined in \cite{Hrapko06}. An improved
method for rotational shear experiments was used, producing an
approximately homogeneous strain field and leading to an enhanced
accuracy. The model was formulated in terms of a large strain
viscoelastic framework and considers nonlinear viscous
deformations in combination with non-linear elastic behavior.

The relative motion of the brain with respect to the skull has
been widely studied to investigate brain injury mechanisms under
impacts, but the motion patterns are not yet thoroughly
understood. The study of \cite{Zou07} analyzed brain motion
patterns using the most recent and advanced experimental relative
brain/skull motion data collected under low-severity impacts. With
a minimum total pseudo-strain energy, the closed-form solutions
for rigid body translation and rotation were obtained by matching
measured neutral density target (NDT) positions with initial NDT
positions. The brain motion was thus separated into rigid body
displacement and deformation. The results showed that the brain
had nearly pure rigid body displacement at low impact speed. As
the impact became more severe, the increased brain motion
primarily was due to deformation, while the rigid body
displacement was limited in magnitude for both translation and
rotation. Under low-severity impacts in the sagittal plane, the
rigid body brain translation had a magnitude of 4-5 mm, and the
whole brain rotation was on the order of +/-5 degrees.

Biomechanical studies using postmortem human subjects (PMHS) in
lateral impact have focused primarily on chest and pelvis
injuries, mechanisms, tolerances, and comparison with side impact
dummies. The objective of \cite{Yoga08} was to determine lateral
impact-induced 3D temporal forces and moments at the head-neck
junction and cranial linear and angular accelerations from sled
tests using PMHS and compare with responses obtained from an
anthropomorphic test device (dummy) designed for lateral impact.
Results indicated that profiles of forces and moments at the
head-neck junction and cranial accelerations were similar between
the two models. However, peak forces and moments at the head-neck
junction, as well as peak cranial linear and angular
accelerations, were lower in the dummy than PMHS. Peak cranial
angular accelerations were suggestive of mild TBI with potential
for loss of consciousness.

Deformation of the human brain was measured \emph{in vivo} by
\cite{Sabet08} in tagged magnetic resonance images (MRI) obtained
dynamically during angular acceleration of the head, in order to
provide quantitative experimental data to illuminate the mechanics
of TBI. Mild angular acceleration was imparted to the skull of a
human volunteer inside an MR scanner, using a custom MR-compatible
device to constrain motion. Deformation of the brain was
characterized quantitatively via Lagrangian strain. Consistent
patterns of radial-circumferential shear strain occurred in the
brain, similar to those observed in models of a viscoelastic gel
cylinder subjected to angular acceleration. It has been noted,
however, that strain fields in the brain are clearly mediated by
the effects of heterogeneity, divisions between regions of the
brain (such as the central fissure and central sulcus) and brain's tethering and suspension system, including the dura mater,
falx cerebri, and tentorium membranes.\bigbreak

The present paper proposes a new approach to brain injury dynamics, phrased as a \emph{coupled loading--rate
hypothesis} for TBI, stating that the main cause of TBI is an
\textsl{external Euclidean jolt}, symbolically an $SE(3)-$jolt, an
impulsive loading striking the head in several
degrees-of-freedom (both translational and rotational) \emph{combined}.
This new concept is radically different from any of the standard FEM techniques proposed so far for brain injury mechanics (see the above literature review), as well as from any kind of Newton--Eulerian or Lagrangian/Hamiltonian injury dynamics, emphasizing 3 new aspects of brain injury mechanics: (i) coupling of all 6 degrees-of-freedom; (ii) jolt dynamics rather than the force dynamics; and (iii) coupled dislocations/disclinations in the Cosserat multipolar viscoelastic continuum brain model.
This new concept is a derivation of our previously defined concept
of the \emph{covariant force law} \cite{13,14,17}.
To support this hypothesis, we develop the coupled Newton--Euler
dynamics of the brains's micro-motions within the cerebrospinal
fluid (see Figure \ref{IvBrain999}), and from it derive the
$SE(3)-$jolt dynamics, as well as its biophysical consequences in
the form of brain's dislocations and disclinations.

\section{The $SE(3)-$jolt: the cause of TBI}

In the language of modern dynamics \cite{13,14,14a,17}, the
microscopic motion of human brain within the skull is governed by
the Euclidean SE(3)--group of 3D motions (see next subsection).
Within brain's SE(3)--group we have both SE(3)--kinematics
(consisting of SE(3)--velocity and its two time derivatives:
SE(3)--acceleration and SE(3)--jerk) and SE(3)--dynamics
(consisting of SE(3)--momentum and its two time derivatives:
SE(3)--force and SE(3)--jolt), which is brain's kinematics
$\times $ brain's mass--inertia distribution.

Informally, the external SE(3)--jolt\footnote{The mechanical
SE(3)--jolt concept is based on the mathematical concept of
higher--order tangency (rigorously defined in terms of jet bundles
of the head's configuration manifold) \cite{14,17}, as follows:
When something hits the human head, or the head hits some external
body, we have a collision. This is naturally described by the
SE(3)--momentum, which is a nonlinear coupling of 3 linear
Newtonian momenta with 3 angular Eulerian momenta. The tangent to
the SE(3)--momentum, defined by the (absolute) time derivative, is
the SE(3)--force. The second-order tangency is given by the
SE(3)--jolt, which is the tangent to the SE(3)--force, also
defined by the time derivative.} is a sharp and sudden change in
the SE(3)--force acting on brain's mass--inertia distribution
(given by brain's mass and inertia matrices). That is, a
`delta'--change in a 3D force--vector coupled to a 3D
torque--vector, striking the head--shell with the brain immersed
into the cerebrospinal fluid. In other words, the SE(3)--jolt is a
sudden, sharp and discontinues shock in all 6 coupled dimensions
of brain's continuous micro--motion within the cerebrospinal
fluid (Figure \ref{IvBrain999}), namely within the three Cartesian
($x,y,z$)--translations and the three corresponding Euler angles
around the Cartesian axes: roll, pitch and yaw. If the SE(3)--jolt
produces a mild shock to the brain (e.g., strong head shake), it
causes mild TBI, with temporary disabled associated sensory-motor
and/or cognitive functions and affecting respiration and movement.
If the SE(3)--jolt produces a hard shock (hitting the head with
external mass), it causes severe TBI, with the total loss of
gesture, speech and movement.

The SE(3)--jolt is rigorously defined in terms of differential
geometry \cite{14,17,20}. Briefly, it is the absolute
time--derivative of the covariant force 1--form (or, co-vector
field). The fundamental law of biomechanics is the \emph{covariant
force law} \cite{13,14,17}, which states:
$$\text{Force co-vector field}=\text{Mass distribution}\times \text{%
Acceleration vector--field},$$ which is formally written (using the Einstein
summation convention, with indices labelling the three Cartesian
translations and the three corresponding Euler angles):
\begin{equation*}
F_{{\mu}}=m_{{\mu}{\nu}}a^{{\nu}},\qquad ({\mu,\nu}%
=1,...,6)
\end{equation*}
where $F_{{\mu}}$ denotes the 6 covariant components of the external
``pushing''\ SE(3)--force co-vector field, $m_{{\mu}{\nu}}$
represents the 6$\times $6 covariant components of brain's
inertia--metric tensor, while $a^{{\nu}}$ corresponds to the 6
contravariant components of brain's internal SE(3)--acceleration
vector-field.

Now, the covariant (absolute, Bianchi) time-derivative
$\frac{{D}}{dt}(\cdot )$ of the covariant SE(3)--force $F_{{\mu
}}$ defines the corresponding external ``striking" SE(3)--jolt
co-vector field:
\begin{equation}
\frac{{D}}{dt}(F_{{\mu }})=m_{{\mu }{\nu }}\frac{{D}}{dt}(a^{{\nu }})=m_{{%
\mu }{\nu }}\left( \dot{a}^{{\nu }}+\Gamma _{\mu \lambda }^{{\nu }}a^{{\mu }%
}a^{{\lambda }}\right) ,  \label{Bianchi}
\end{equation}%
where ${\frac{{D}}{dt}}{(}a^{{\nu }})$ denotes the 6 contravariant
components of brain's internal SE(3)--jerk vector-field and
overdot ($\dot{~}$) denotes the time derivative. $\Gamma _{\mu
\lambda }^{{\nu }}$ are the Christoffel's symbols of the
Levi--Civita connection for the SE(3)--group, which are zero in
case of pure Cartesian translations and nonzero in case of
rotations as well as in the full--coupling of translations and
rotations.

In the following, we elaborate on the SE(3)--jolt concept (using
vector and tensor methods) and its biophysical TBI consequences in
the form of brain's dislocations and disclinations.

\subsection{$SE(3)-$group of brain's micro--motions within the CSF}

The brain and the CSF together exhibit periodic microscopic
translational and rotational motion in a pulsatile fashion to and
from the cranial cavity, in the frequency range of normal heart
rate (with associated periodic squeezing of brain's
ventricles) \cite{Maier94}. This micro--motion is mathematically
defined by the Euclidean (gauge) $SE(3)-$group. Briefly, the
$SE(3)-$group is
defined as a semidirect (noncommutative) product of 3D rotations and
3D translations, $$SE(3):=SO(3)\rhd \Bbb{R}^{3}.$$ Its most important subgroups are the
following (see Appendix for technical details):
\begin{center}
{{\frame{$
\begin{array}{cc}
\mathbf{Subgroup} & \mathbf{Definition} \\ \hline
\begin{array}{c}
SO(3),\text{ group of rotations} \\
\text{in 3D (a spherical joint)}
\end{array}
&
\begin{array}{c}
\text{Set of all proper orthogonal } \\
3\times 3-\text{rotational matrices}
\end{array}
\\ \hline
\begin{array}{c}
SE(2),\text{ special Euclidean group} \\
\text{in 2D (all planar motions)}
\end{array}
&
\begin{array}{c}
\text{Set of all }3\times 3-\text{matrices:} \\
\left[
\begin{array}{ccc}
\cos \theta & \sin \theta & r_{x} \\
-\sin \theta & \cos \theta & r_{y} \\
0 & 0 & 1
\end{array}
\right]
\end{array}
\\ \hline
\begin{array}{c}
SO(2),\text{ group of rotations in 2D} \\
\text{subgroup of }SE(2)\text{--group} \\
\text{(a revolute joint)}
\end{array}
&
\begin{array}{c}
\text{Set of all proper orthogonal } \\
2\times 2-\text{rotational matrices} \\
\text{ included in }SE(2)-\text{group}
\end{array}
\\ \hline
\begin{array}{c}
\Bbb{R}^{3},\text{ group of translations in 3D} \\
\text{(all spatial displacements)}
\end{array}
& \text{Euclidean 3D vector space}
\end{array}
$}}}\end{center}

In other words, the gauge $SE(3)-$group of Euclidean micro-motions
of the brain immersed in the cerebrospinal fluid within the
cranial cavity, contains matrices of the form {\small $ \left(
\begin{array}{cc}
{\bf R} & {\bf b} \\
0 & 1
\end{array}
\right), $} where ${\bf b}$ is brain's 3D micro-translation
vector and ${\bf R}$ is brain's 3D rotation matrix, given by
the product ${\bf R}=R_{\varphi }\cdot R_{\psi }\cdot R_{\theta }$
of brain's three Eulerian micro-rotations,
$\text{roll}=R_{\varphi },~\text{pitch}=R_{\psi
},~\text{yaw}=R_{\theta }$,
performed respectively about the $x-$axis by an angle $%
\varphi ,$ about the $y-$axis by an angle $\psi ,$ and about the
$z-$axis by an angle $\theta $ \cite{VladSIAM,ParkChung,IJHR},
{\small
\begin{equation*}
R_{\varphi } =\left[
\begin{array}{ccc}
1 & 0 & 0 \\
0 & \cos \varphi & -\sin \varphi \\
0 & \sin \varphi & \cos \varphi%
\end{array}
\right] , ~~ R_{\psi } =\left[
\begin{array}{ccc}
\cos \psi & 0 & \sin \psi \\
0 & 1 & 0 \\
-\sin \psi & 0 & \cos \psi%
\end{array}
\right] , ~~ R_{\theta } =\left[
\begin{array}{ccc}
\cos \theta & -\sin \theta & 0 \\
\sin \theta & \cos \theta & 0 \\
0 & 0 & 1%
\end{array}
\right].
\end{equation*}}

Therefore, brain's natural $SE(3)-$dynamics within the
cerebrospinal fluid is given by the coupling of Newtonian
(translational) and Eulerian (rotational) equations of
micro-motion.

\subsection{Brain's natural $SE(3)-$dynamics}

To support our coupled loading--rate hypothesis, we formulate the coupled
Newton--Euler dynamics of brain's micro-motions within the scull's $%
SE(3)-$group of motions. The forced Newton--Euler equations read in vector
(boldface) form
\begin{eqnarray}
\text{Newton} &:&~\mathbf{\dot{p}}~\mathbf{\equiv M\dot{v}=F+p\times \omega }%
,  \label{vecForm} \\
\text{Euler} &:&~\mathbf{\dot{\pi}}~\mathbf{\equiv I\dot{\omega}=T+\pi
\times \omega +p\times v},  \notag
\end{eqnarray}
where $\times $ denotes
the vector cross product,\footnote{%
Recall that the cross product $\mathbf{u\times v}$ of two vectors $\mathbf{u}
$ and $\mathbf{v}$ equals $\mathbf{u\times v}=uv\func{sin}\theta \mathbf{n}$%
, where $\theta $ is the angle between $\mathbf{u}$ and $\mathbf{v}$, while $%
\mathbf{n}$ is a unit vector perpendicular to the plane of $\mathbf{u}$ and $%
\mathbf{v}$ such that $\mathbf{u}$ and $\mathbf{v}$ form a right-handed
system.}
\begin{equation*}
\mathbf{M}\equiv M_{ij}=diag\{m_{1},m_{2},m_{3}\}\qquad \text{and}\qquad
\mathbf{I}\equiv I_{ij}=diag\{I_{1},I_{2},I_{3}\},\qquad(i,j=1,2,3)
\end{equation*}
are brain's (diagonal) mass and inertia matrices,\footnote{%
In reality, mass and inertia matrices ($\mathbf{M,I}$) are not diagonal but
rather full $3\times 3$ positive--definite symmetric matrices with coupled
mass-- and inertia--products. Even more realistic, fully--coupled
mass--inertial properties of a brain immersed in (incompressible,
irrotational and inviscid) cerebrospinal fluid are defined by the single
non-diagonal $6\times 6$ positive--definite symmetric mass--inertia matrix $%
\mathcal{M}_{SE(3)}$, the so-called material metric tensor of the $SE(3)-$%
group, which has all nonzero mass--inertia coupling products. In other
words, the $6\times 6$ matrix $\mathcal{M}_{SE(3)}$ contains: (i) brain's own mass plus the added mass matrix associated with the fluid, (ii)
brain's own inertia plus the added inertia matrix associated with the
potential flow of the fluid, and (iii) all the coupling terms between linear
and angular momenta. However, for simplicity, in this paper we shall
consider only the simple case of two separate diagonal $3\times 3$ matrices (%
$\mathbf{M,I}$).} defining brain's mass--inertia distribution, with
principal inertia moments given in Cartesian coordinates ($x,y,z$) by volume
integrals
\begin{equation*}
I_{1}=\iiint \rho (z^{2}+y^{2})dxdydz,~~I_{2}=\iiint \rho
(x^{2}+z^{2})dxdydz,~~I_{3}=\iiint \rho (x^{2}+y^{2})dxdydz,
\end{equation*}
dependent on brain's density $\rho =\rho (x,y,z)$,
\begin{equation*}
\mathbf{v}\equiv v^{i}=[v_{1},v_{2},v_{3}]^{t}\qquad \text{and\qquad }%
\mathbf{\omega }\equiv {\omega }^{i}=[\omega _{1},\omega _{2},\omega
_{3}]^{t}
\end{equation*}
(where $[~]^{t}$ denotes the vector transpose) are brain's linear and
angular velocity vectors\footnote{%
In reality, $\mathbf{\omega }$ is a $3\times 3$ \emph{attitude matrix} (see
Appendix). However, for simplicity, we will stick to the (mostly)
symmetrical translation--rotation vector form.} (that is, column vectors),
\begin{equation*}
\mathbf{F}\equiv F_{i}=[F_{1},F_{2},F_{3}]\qquad \text{and}\qquad \mathbf{T}%
\equiv T_{i}=[T_{1},T_{2},T_{3}]
\end{equation*}
are gravitational and other external force and torque co-vectors (that is,
row vectors) acting on the brain within the scull,
\begin{eqnarray*}
\mathbf{p} &\equiv &p_{i}\equiv \mathbf{Mv}%
=[p_{1},p_{2},p_{3}]=[m_{1}v_{1},m_{2}v_{2},m_{2}v_{2}]\qquad \text{and} \\
\mathbf{\pi } &\equiv &\pi _{i}\equiv \mathbf{I\omega }=[\pi _{1},\pi
_{2},\pi _{3}]=[I_{1}\omega _{1},I_{2}\omega _{2},I_{3}\omega _{3}]
\end{eqnarray*}
are brain's linear and angular momentum co-vectors.

In tensor form, the forced Newton--Euler equations (\ref{vecForm}) read
\begin{eqnarray*}
\dot{p}_{i} &\equiv &M_{ij}\dot{v}^{j}=F_{i}+\varepsilon _{ik}^{j}p_{j}{%
\omega }^{k},\qquad(i,j,k=1,2,3) \\
\dot{\pi}_{i} &\equiv &I_{ij}\dot{\omega}^{j}=T_{i}+\varepsilon _{ik}^{j}\pi
_{j}\omega ^{k}+\varepsilon _{ik}^{j}p_{j}v^{k},
\end{eqnarray*}
where the permutation symbol $\varepsilon _{ik}^{j}$ is\ defined as
\begin{equation*}
\varepsilon _{ik}^{j}=
\begin{cases}
+1 & \text{if }(i,j,k)\text{ is }(1,2,3),(3,1,2)\text{ or }(2,3,1), \\
-1 & \text{if }(i,j,k)\text{ is }(3,2,1),(1,3,2)\text{ or }(2,1,3), \\
0 & \text{otherwise: }i=j\text{ or }j=k\text{ or }k=i.
\end{cases}
\end{equation*}

In scalar form, the forced Newton--Euler equations (\ref{vecForm}) expand as
\begin{eqnarray}
\text{Newton} &:&\left\{
\begin{array}{c}
\dot{p}_{_{1}}={F_{1}}-{m_{3}}{v_{3}}{\omega _{2}}+{m_{2}}{v_{2}}{\omega _{3}%
} \\
\dot{p}_{_{2}}={F_{2}}+{m_{3}}{v_{3}}{\omega _{1}}-{m_{1}}{v_{1}}{\omega _{3}%
} \\
\dot{p}_{_{3}}={F_{3}}-{m_{2}}{v_{2}}{\omega _{1}}+{m_{1}}{v_{1}}{\omega _{2}%
}
\end{array}
\right. ,  \label{scalarForm} \\
\text{Euler} &:&\left\{
\begin{array}{c}
\dot{\pi}_{_{1}}={T_{1}}+({m_{2}}-{m_{3}}){v_{2}}{v_{3}}+({I_{2}}-{I_{3}}){%
\omega _{2}}{\omega _{3}} \\
\dot{\pi}_{_{2}}={T_{2}}+({m_{3}}-{m_{1}}){v_{1}}{v_{3}}+({I_{3}}-{I_{1}}){%
\omega _{1}}{\omega _{3}} \\
\dot{\pi}_{_{3}}={T_{3}}+({m_{1}}-{m_{2}}){v_{1}}{v_{2}}+({I_{1}}-{I_{2}}){%
\omega _{1}}{\omega _{2}}
\end{array}
\right. ,  \notag
\end{eqnarray}
showing brain's individual mass and inertia couplings.

Equations (\ref{vecForm})--(\ref{scalarForm}) can be derived from the
translational + rotational kinetic energy of the brain\footnote{%
In a fully--coupled Newton--Euler brain dynamics, instead of equation (\ref
{Ek}) we would have brain's kinetic energy defined by the inner product:
\begin{equation*}
E_{k}=\frac{1}{2}\left[ \QOVERD( ) {\mathbf{p}}{\mathbf{\pi }}\left|
\mathcal{M}_{SE(3)}\right. \QOVERD( ) {\mathbf{p}}{\mathbf{\pi }}\right] .
\end{equation*}
}
\begin{equation}
E_{k}={\frac{1}{2}}\mathbf{v}^{t}\mathbf{Mv}+{\frac{1}{2}}\mathbf{\omega }%
^{t}\mathbf{I\omega },  \label{Ek}
\end{equation}
or, in tensor form
\begin{equation*}
E={\frac{1}{2}}M_{ij}{v}^{i}{v}^{j}+{\frac{1}{2}}I_{ij}{\omega}%
^{i}{\omega}^{j}.
\end{equation*}

For this we use the \emph{Kirchhoff--Lagrangian equations} (see, e.g., \cite
{Kirchhoff,naomi97}, or the original work of Kirchhoff in German)
\begin{eqnarray}
\frac{d}{{dt}}\partial _{\mathbf{v}}E_{k} &=&\partial _{\mathbf{v}%
}E_{k}\times \mathbf{\omega }+\mathbf{F},  \label{Kirch} \\
{\frac{d}{{dt}}}\partial _{\mathbf{\omega }}E_{k} &=&\partial _{\mathbf{%
\omega }}E_{k}\times \mathbf{\omega }+\partial _{\mathbf{v}}E_{k}\times
\mathbf{v}+\mathbf{T},  \notag
\end{eqnarray}
where $\partial _{\mathbf{v}}E_{k}=\frac{\partial E_{k}}{\partial \mathbf{v}}%
,~\partial _{\mathbf{\omega }}E_{k}=\frac{\partial E_{k}}{\partial \mathbf{%
\omega }}$; in tensor form these equations read
\begin{eqnarray*}
\frac{d}{dt}\partial _{v^{i}}E &=&\varepsilon _{ik}^{j}\left( \partial
_{v^{j}}E\right) \omega ^{k}+F_{i}, \\
\frac{d}{dt}\partial _{{\omega }^{i}}E &=&\varepsilon _{ik}^{j}\left(
\partial _{{\omega }^{j}}E\right) {\omega }^{k}+\varepsilon _{ik}^{j}\left(
\partial _{v^{j}}E\right) v^{k}+T_{i}.
\end{eqnarray*}

Using (\ref{Ek})--(\ref{Kirch}), brain's linear and angular momentum
co-vectors are defined as
\begin{equation*}
\mathbf{p}=\partial _{\mathbf{v}}E_{k}{,\qquad \mathbf{\pi }=\partial _{%
\mathbf{\omega }}E_{k},}
\end{equation*}
or, in tensor form
\begin{equation*}
p_{i}=\partial _{v^{i}}E{,\qquad }\pi _{i}=\partial _{{\omega }^{i}}E,
\end{equation*}
with their corresponding time derivatives, in vector form
\begin{equation*}
~\mathbf{\dot{p}}=\frac{d}{dt}\mathbf{p=}\frac{d}{dt}\partial _{\mathbf{v}}E{%
,\qquad \mathbf{\dot{\pi}}=}\frac{d}{dt}\mathbf{\pi =}\frac{d}{dt}\partial _{%
\mathbf{\omega }}E,
\end{equation*}
or, in tensor form
\begin{equation*}
~\dot{p}_{i}=\frac{d}{dt}p_{i}=\frac{d}{dt}\partial _{v^{i}}E{,\qquad \dot{%
\pi}_{i}=}\frac{d}{dt}\pi _{i}=\frac{d}{dt}\partial _{{\omega }^{i}}E,
\end{equation*}
or, in scalar form
\begin{equation*}
\mathbf{\dot{p}}=[\dot{p}_{1},\dot{p}_{2},\dot{p}_{3}]=[m_{1}\dot{v}%
_{1},m_{2}\dot{v}_{2},m_{3}\dot{v}_{3}],\qquad {\mathbf{\dot{\pi}}}=[\dot{\pi%
}_{1},\dot{\pi}_{2},\dot{\pi}_{3}]=[I_{1}\dot{\omega}_{1},I_{2}\dot{\omega}%
_{2},I_{3}\dot{\omega}_{3}].
\end{equation*}

While brain's healthy $SE(3)-$dynamics within the cerebrospinal fluid is
given by the coupled Newton--Euler micro--dynamics, the TBI is actually
caused by the sharp and discontinuous change in this natural $SE(3)$
micro-dynamics, in the form of the $SE(3)-$jolt, causing brain's
discontinuous deformations.

\subsection{Brain's traumatic dynamics: the $SE(3)-$jolt}

The $SE(3)-$jolt, the actual cause of the TBI (in the form of the brain's
plastic deformations), is defined as a coupled Newton+Euler jolt; in
(co)vector form the $SE(3)-$jolt reads\footnote{%
Note that the derivative of the cross--product of two vectors follows the
standard calculus product--rule: $\frac{d}{dt}(\mathbf{u\times v})=\mathbf{%
\dot{u}\times v+u\times \dot{v}.}$}
\begin{equation*}
SE(3)-\text{jolt}:\left\{
\begin{array}{l}
\text{Newton~jolt}:\mathbf{\dot{F}=\ddot{p}-\dot{p}\times \omega -p\times
\dot{\omega}}~,\qquad \\
\text{Euler~jolt}:\mathbf{\dot{T}=\ddot{\pi}}~\mathbf{-\dot{\pi}\times
\omega -\pi \times \dot{\omega}-\dot{p}\times v-p\times \dot{v}},
\end{array}
\right.
\end{equation*}
where the linear and angular jolt co-vectors are
\begin{equation*}
\mathbf{\dot{F}\equiv M\ddot{v}}=[\dot{F}_{{1}},\dot{F}_{{2}},\dot{F}_{{3}%
}],\qquad \mathbf{\dot{T}\equiv I\ddot{\omega}}=[\dot{T}_{{1}},\dot{T}_{{2}},%
\dot{T}_{{3}}],
\end{equation*}
where
\begin{equation*}
\mathbf{\ddot{v}}=[\ddot{v}_{{1}},\ddot{v}_{{2}},\ddot{v}_{{3}}]^{t},\qquad
\mathbf{\ddot{\omega}}=[\ddot{\omega}_{{1}},\ddot{\omega}_{{2}},\ddot{\omega}%
_{{3}}]^{t},
\end{equation*}
are linear and angular jerk vectors.

In tensor form, the $SE(3)-$jolt reads\footnote{%
In this paragraph the overdots actually denote the absolute
Bianchi (covariant) time-derivative (\ref{Bianchi}), so that the
jolts retain the proper covector character, which would be lost if
ordinary time derivatives are used. However, for the sake of
simplicity and wider readability, we stick to the same overdot
notation.}
\begin{eqnarray*}
~\dot{F}_{i} &=&\ddot{p}_{i}-\varepsilon _{ik}^{j}\dot{p}_{j}{\omega }%
^{k}-\varepsilon _{ik}^{j}p_{j}{\dot{\omega}}^{k}, \qquad(i,j,k=1,2,3) \\
~\dot{T}_{{i}} &=&\ddot{\pi}_{i}~-\varepsilon _{ik}^{j}\dot{\pi}_{j}\omega
^{k}-\varepsilon _{ik}^{j}\pi _{j}{\dot{\omega}}^{k}-\varepsilon _{ik}^{j}%
\dot{p}_{j}v^{k}-\varepsilon _{ik}^{j}p_{j}\dot{v}^{k},
\end{eqnarray*}
in which the linear and angular jolt covectors are defined as
\begin{eqnarray*}
\mathbf{\dot{F}} &\equiv &\dot{F}_{i}=\mathbf{M\ddot{v}}\,\equiv \mathbf{\,}%
M_{ij}\ddot{v}^{j}=[\dot{F}_{1},\dot{F}_{2},\dot{F}_{3}], \\
\mathbf{\dot{T}} &\equiv &\dot{T}_{{i}}=\mathbf{I\ddot{\omega}\equiv \,}%
I_{ij}\ddot{\omega}^{j}=[\dot{T}_{{1}},\dot{T}_{{2}},\dot{T}_{{3}}],
\end{eqnarray*}
where \ $\mathbf{\ddot{v}}=\ddot{v}^{{i}},$ and $\mathbf{\ddot{\omega}}=%
\ddot{\omega}^{{i}}$ are linear and angular jerk vectors.

In scalar form, the $SE(3)-$jolt expands as
\begin{eqnarray*}
\text{Newton~jolt} &:&\left\{
\begin{array}{l}
\dot{F}_{{1}}=\ddot{p}_{1}-m_{{2}}\omega _{{3}}\dot{v}_{{2}}+m_{{3}}\left( {%
\omega }_{{2}}\dot{v}_{{3}}+v_{{3}}\dot{\omega}_{{2}}\right) -m_{{2}}v_{{2}}{%
\dot{\omega}}_{{3}}, \\
\dot{F}_{{2}}=\ddot{p}_{2}+m_{{1}}\omega _{{3}}\dot{v}_{{1}}-m_{{3}}\omega _{%
{1}}\dot{v}_{{3}}-m_{{3}}v_{{3}}\dot{\omega}_{{1}}+m_{{1}}v_{{1}}\dot{\omega}%
_{{3}}, \\
\dot{F}_{{3}}=\ddot{p}_{3}-m_{{1}}\omega _{{2}}\dot{v}_{{1}}+m_{{2}}\omega _{%
{1}}\dot{v}_{{2}}-v_{{2}}\dot{\omega}_{{1}}-m_{{1}}v_{{1}}\dot{\omega}_{{2}},
\end{array}
\right. \\
&& \\
\text{Euler~jolt} &:&\left\{
\begin{array}{l}
\dot{T}_{{1}}=\ddot{\pi}_{1}-(m_{{2}}-m_{{3}})\left( v_{{3}}\dot{v}_{{2}}+v_{%
{2}}\dot{v}_{{3}}\right) -(I_{{2}}-I_{{3}})\left( \omega _{{3}}\dot{\omega}_{%
{2}}+{\omega }_{{2}}{\dot{\omega}}_{{3}}\right) , \\
\dot{T}_{{2}}=\ddot{\pi}_{2}+(m_{{1}}-m_{{3}})\left( v_{{3}}\dot{v}_{{1}}+v_{%
{1}}\dot{v}_{{3}}\right) +(I_{{1}}-I_{{3}})\left( {\omega }_{{3}}{\dot{\omega%
}}_{{1}}+{\omega }_{{1}}{\dot{\omega}}_{{3}}\right) , \\
\dot{T}_{{3}}=\ddot{\pi}_{3}-(m_{{1}}-m_{{2}})\left( v_{{2}}\dot{v}_{{1}}+v_{%
{1}}\dot{v}_{{2}}\right) -(I_{{1}}-I_{{2}})\left( {\omega }_{{2}}{\dot{\omega%
}}_{{1}}+{\omega }_{{1}}{\dot{\omega}}_{{2}}\right).
\end{array}
\right.
\end{eqnarray*}

We remark here that the linear and angular momenta ($\mathbf{p,\pi
}$), forces ($\mathbf{F,T}$) and jolts ($\mathbf{\dot{F},\dot{T}}$) are
co-vectors (row vectors), while the linear and angular velocities ($\mathbf{%
v,\omega }$), accelerations ($\mathbf{\dot{v},\dot{\omega}}$) and jerks ($%
\mathbf{\ddot{v},\ddot{\omega}}$) are vectors (column vectors).
This bio-physically means that the `jerk' vector should not be
confused with the `jolt' co-vector. For example, the `jerk'\ means
shaking the head's own mass--inertia matrices (mainly in the
atlanto--occipital and atlanto--axial joints), while the
`jolt'means actually hitting the head with some external
mass--inertia matrices included in the `hitting'\ SE(3)--jolt, or
hitting some external static/massive body with the head (e.g., the
ground -- gravitational effect, or the wall -- inertial effect).
Consequently, the mass-less `jerk' vector\ represents a (translational+rotational) \textit{%
non-collision effect} that can cause only weaker brain injuries, while the
inertial `jolt'\ co-vector represents a (translational+rotational) \textit{%
collision effect} that can cause hard brain injuries.

For example, while driving a car, the SE(3)--jerk of the
head--neck system happens every time the driver brakes abruptly.
On the other hand, the SE(3)--jolt means actual impact to the
head. Similarly, the whiplash--jerk, caused by rear--end car
collisions, is like a soft version of the high pitch--jolt caused
by the boxing `upper-cut'. Also, violently shaking the head
left--right in the transverse plane is like a soft version of the
high yaw--jolt caused by the boxing `cross-cut'.

\subsection{Brain's dislocations and disclinations caused by the $SE(3)-$%
jolt}

Recall from introduction that for mild TBI, the best injury predictor
is considered to be the product of brain's strain and strain rate, which is the standard
isotropic viscoelastic continuum concept. To improve this standard concept,
in this subsection, we consider human brain as a 3D anisotropic
multipolar \emph{Cosserat viscoelastic continuum}
\cite{Cosserat1,Cosserat2,Eringen02}, exhibiting
coupled--stress--strain elastic properties. This non-standard
continuum model is suitable for analyzing plastic (irreversible)
deformations and fracture mechanics \cite{Bilby} in multi-layered
materials with microstructure (in which slips and bending of
layers introduces additional degrees of freedom, non-existent in
the standard continuum models; see \cite{Mindlin65,Lakes85} for
physical characteristics and \cite {Yang81,Yang82,Park86} for
biomechanical applications).

The $SE(3)-$jolt $(\mathbf{\dot{F},\dot{T}})$ causes two types of
brain's rapid discontinuous deformations:

\begin{enumerate}
\item  The Newton jolt $\mathbf{\dot{F}}$ can cause
micro-translational \emph{dislocations}, or discontinuities in the
Cosserat translations;

\item  The Euler jolt $\mathbf{\dot{T}}$ can cause micro-rotational \emph{%
disclinations}, or discontinuities in the Cosserat rotations.
\end{enumerate}

For general treatment on dislocations and disclinations related to
asymmetric discontinuous deformations in multipolar materials,
see, e.g., \cite{Jian95,Yang01}.

To precisely define brain's dislocations and disclinations,
caused by the $SE(3)-$jolt $(\mathbf{\dot{F},\dot{T}})$, we first
define the coordinate co-frame, i.e., the set of basis 1--forms
$\{dx^{i}\}$, given in local coordinates
$x^{i}=(x^{1},x^{2},x^{3})=(x,y,z)$, attached to brain's
center-of-mass. Then, in the coordinate co-frame $\{dx^{i}\}$ we
introduce the following set of brain's plastic--deformation--related $%
SE(3)-$based differential $p-$forms\footnote{%
Differential $p-$forms are totally skew-symmetric covariant
tensors, defined using the exterior wedge--product and exterior
derivative. The proper definition of exterior derivative $d$ for a
$p-$form $\beta $ on a smooth
manifold $M$, includes the \textit{Poincar\'{e} lemma} \cite{14,17}: {$%
d(d\beta )=0$}, and validates the \textit{general Stokes formula}
\[
\int_{\partial M}\beta =\int_{M}d\beta ,
\]
where $M$ is a $p-$dimensional \emph{manifold with a boundary} and
$\partial M$ is its $(p-1)-$dimensional \emph{boundary}, while the
integrals have appropriate dimensions.
\par
A $p-$form $\beta $ is called \emph{closed} if its exterior
derivative is equal to zero,
\[
d\beta =0.
\]
From this condition one can see that the closed form (the
\emph{kernel} of the exterior derivative operator $d$) is
conserved quantity. Therefore, closed $p-$forms possess certain
invariant properties, physically corresponding to the conservation
laws.
\par
A $p-$form $\beta $ that is an exterior derivative of some $(p-1)-$form $%
\alpha $,
\[
\beta =d\alpha ,
\]
is called \emph{exact} (the \emph{image} of the exterior
derivative operator $d$). By \textit{Poincar\'{e} lemma}, exact
forms prove to be closed automatically,
\[
d\beta =d(d\alpha )=0.
\]
This lemma is the foundation of the de Rham cohomology theory \cite{14,17,20}.} (see, e.g., \cite{14,17}):\newline\\
$~~~~$the \emph{dislocation current }1--form, $\mathbf{J}=J_{i}\,dx^{i};$%
\newline
$~~~~$the \emph{dislocation density }2--form, $\mathbf{\alpha }=\frac{1}{2}%
\alpha _{ij}\,dx^{i}\wedge dx^{j};$\newline
$~~~~$the \emph{disclination current }2--form, $\mathbf{S}=\frac{1}{2}%
S_{ij}\,dx^{i}\wedge dx^{j};$ ~and\newline
$~~~~$the \emph{disclination density }3--form, $\mathbf{Q}=\frac{1}{3!}%
Q_{ijk}\,dx^{i}\wedge dx^{j}\wedge dx^{k}$,

where $\wedge $ denotes the exterior wedge--product. According to
Edelen \cite{Edelen,Kadic}, these four $SE(3)-$based differential forms
satisfy the following set of continuity equations:
\begin{eqnarray}
&&\mathbf{\dot{\alpha}}=\mathbf{-dJ-S,}  \label{dis1} \\
&&\mathbf{\dot{Q}}=\mathbf{-dS,}  \label{dis2} \\
&&\mathbf{d\alpha }=\mathbf{Q,}  \label{dis3} \\
&&\mathbf{dQ}=\mathbf{0,}\qquad   \label{dis4}
\end{eqnarray}
where $\mathbf{d}$ denotes the exterior derivative.

In components, the simplest, fourth equation (\ref{dis4}),
representing the \emph{Bianchi identity}, can be rewritten as
\[
\mathbf{dQ}=\partial _{l}Q_{[ijk]}\,dx^{l}\wedge dx^{i}\wedge
dx^{j}\wedge dx^{k}=0,
\]
where $\partial _{i}\equiv\partial /\partial x^{i}$, while $\theta
_{\lbrack ij...]}$ denotes the skew-symmetric part of $\theta
_{ij...}$.

Similarly, the third equation (\ref{dis3}) in components reads
\begin{eqnarray*}
\frac{1}{3!}Q_{ijk}\,dx^{i}\wedge dx^{j}\wedge dx^{k} &=&\partial
_{k}\alpha
_{\lbrack ij]}\,dx^{k}\wedge dx^{i}\wedge dx^{j},\text{\qquad or} \\
Q_{ijk} &=&-6\partial _{k}\alpha _{\lbrack ij]}.
\end{eqnarray*}

The second equation (\ref{dis2}) in components reads
\begin{eqnarray*}
\frac{1}{3!}\dot{Q}_{ijk}\,dx^{i}\wedge dx^{j}\wedge dx^{k}
&=&-\partial
_{k}S_{[ij]}\,dx^{k}\wedge dx^{i}\wedge dx^{j},\text{\qquad or} \\
\dot{Q}_{ijk} &=&6\partial _{k}S_{[ij]}.
\end{eqnarray*}

Finally, the first equation (\ref{dis1}) in components reads
\begin{eqnarray*}
\frac{1}{2}\dot{\alpha}_{ij}\,dx^{i}\wedge dx^{j} &=&(\partial _{j}J_{i}-%
\frac{1}{2}S_{ij})\,dx^{i}\wedge dx^{j},\text{\qquad or} \\
\dot{\alpha}_{ij}\, &=&2\partial _{j}J_{i}-S_{ij}\,.
\end{eqnarray*}

In words, we have:

\begin{itemize}
\item  The 2--form equation (\ref{dis1}) defines the time derivative $%
\mathbf{\dot{\alpha}=}\frac{1}{2}\dot{\alpha}_{ij}\,dx^{i}\wedge
dx^{j}$ of the dislocation density $\mathbf{\alpha }$ as the
(negative) sum of the
disclination current $\mathbf{S}$ and the curl of the dislocation current $%
\mathbf{J}$.

\item  The 3--form equation (\ref{dis2}) states that the time derivative $%
\mathbf{\dot{Q}=}\frac{1}{3!}\dot{Q}_{ijk}\,dx^{i}\wedge
dx^{j}\wedge dx^{k}$ of the disclination density $\mathbf{Q}$ is
the (negative) divergence of the disclination current
$\mathbf{S}$.

\item  The 3--form equation (\ref{dis3}) defines the disclination density $%
\mathbf{Q}$ as the divergence of the dislocation density
$\mathbf{\alpha }$, that is, $\mathbf{Q}$ is the \emph{exact}
3--form.

\item  The Bianchi identity (\ref{dis4}) follows from equation
(\ref{dis3})
by \textit{Poincar\'{e} lemma} and states that the disclination density $%
\mathbf{Q}$ is conserved quantity, that is, $\mathbf{Q}$ is the
\emph{closed} 3--form. Also, every 4--form in 3D space is zero.
\end{itemize}

From these equations, we can derive two important conclusions:

\begin{enumerate}
\item  Being the derivatives of the
dislocations, brain's disclinations are higher--order tensors,
and thus more complex quantities, which means that they present
a higher risk for the severe TBI than dislocations --- a fact
which \emph{is} supported by the literature (see review of
existing TBI--models given in Introduction).

\item  Brain's dislocations and disclinations are mutually
coupled by the underlaying $SE(3)-$group, which means that we
cannot separately analyze translational and rotational TBIs --- a
fact which \emph{is not} supported by the literature.
\end{enumerate}

\section{Conclusion}

Based on the previously developed covariant force law,
in this paper we have formulated a new coupled loading--rate hypothesis
for the TBI, which states that the main cause of traumatic brain
injury is an external $SE(3)-$jolt, an impulsive loading striking
the head in several degrees-of-freedom, both rotational and
translational, combined.\footnote{One practical application of the
proposed model is in design of helmets. Briefly, a `hard' helmet
saves the skull but not the brain; alternatively, a `soft' helmet
protects the brain from the collision jolt but does not protect
the skull. A good helmet is both `hard' and `soft'. In other
words, if a human head covered with a solid helmet collides with a
massive external body, the skull will be protected by the helmet
-- but the brain will still be shocked by the SE(3)--jolt, and a
TBI will be caused. With or without the `hard' helmet, brain's
inertia tensor will be moved and rotated by the external
SE(3)--jolt, and this will cause a brain injury, proportional to
the jolt--collision with the head. Therefore, while protecting the
skull is a necessary condition for protecting the brain, it is not
enough. Brain's inertia tensor needs another kind of
protection from the external collision-jolt. Contrastingly, if a
human head covered with a `soft' helmet collides with a massive
external body, the helmet will dissipate the energy from the
collision jolt, but will not necessarily protect the skull. As a
result, a proper helmet would have to have both a hard external
shell (to protect the skull) and a soft internal part (that will
dissipate the energy from the collision jolt by its own
destruction, in the same way as a car saves its passengers from
the collision jolt by its own destruction). Note that,
hypothetically speaking, an ideal shock--absorber is not a
classical spring--damper system (with the distance--dependent
spring and velocity--dependent damper), but rather a
\emph{constant--resistance damper}.} To demonstrate this, we have
developed the vector Newton--Euler mechanics on the Euclidean
$SE(3)-$group of brain's micro-motions within the
cerebrospinal fluid. In this way, we have precisely defined the
concept of the $SE(3)-$jolt, which is a cause of brain's rapid
discontinuous deformations: (i) translational dislocations,
and (ii) rotational disclinations. Based on the presented
model, we argue that: (1) rapid discontinuous rotations present a
higher risk for the severe TBI than rapid discontinuous
translations, and (2) that we cannot separately analyze rapid
brain's rotations from translations, as they are in reality
coupled.

\section{Appendix: The $SE(3)-$group}

Special Euclidean group $SE(3):=SO(3)\rhd \Bbb{R}^{3}$, (the semidirect
product of the group of rotations with the corresponding group of
translations), is the Lie group consisting of isometries of the Euclidean 3D
space $\Bbb{R}^{3}$.

An element of $SE(3)$ is a pair $(A,a)$ where $A\in SO(3)$ and $a\in \Bbb{R}%
^{3}.$ The action of $SE(3)$ on $\Bbb{R}^{3}$ is the rotation $A$ followed
by translation by the vector $a$ and has the expression
\[
(A,a)\cdot x=Ax+a.
\]

The Lie algebra of the Euclidean group $SE(3)$ is $\mathfrak{se}(3)=\Bbb{R}%
^{3}\times \Bbb{R}^{3}$ with the Lie bracket
\begin{equation}
\lbrack (\xi ,u),(\eta ,v)]=(\xi \times \eta ,\xi \times v-\eta \times u).
\label{lbse3}
\end{equation}

Using homogeneous coordinates, we can represent $SE(3)$ as follows,
\[
SE(3)=\ \ \left\{ \left(
\begin{array}{cc}
R & p \\
0 & 1
\end{array}
\right) \in GL(4,\Bbb{R}):R\in SO(3),\,p\in \Bbb{R}^{3}\right\} ,
\]
with the action on $\Bbb{R}^{3}$ given by the usual matrix--vector product
when we identify $\Bbb{R}^{3}$ with the section $\Bbb{R}^{3}\times
\{1\}\subset \Bbb{R}^{4}$. In particular, given
\[
g=\left(
\begin{array}{cc}
R & p \\
0 & 1
\end{array}
\right) \in SE(3),
\]
and $q\in \Bbb{R}^{3}$, we have
\[
g\cdot q=Rq+p,
\]
or as a matrix--vector product,
\[
\left(
\begin{array}{cc}
R & p \\
0 & 1
\end{array}
\right) \left(
\begin{array}{c}
q \\
1
\end{array}
\right) =\left(
\begin{array}{c}
Rq+p \\
1
\end{array}
\right) .
\]

The Lie algebra of $SE(3)$, denoted $\mathfrak{se}(3)$, is given by \
\[
\mathfrak{se}(3)=\ \ \left\{ \left(
\begin{array}{cc}
\omega & v \\
0 & 0
\end{array}
\right) \in M_{4}(\Bbb{R}):\omega\in \mathfrak{so}(3),\,v\in \Bbb{R}%
^{3}\right\} ,
\]
where the attitude (or, angular velocity) matrix $\omega:\Bbb{R}%
^{3}\rightarrow \mathfrak{so}(3)$ is given by
\[
\omega=\left(
\begin{array}{ccc}
0 & -\omega _{z} & \omega _{y} \\
\omega _{z} & 0 & -\omega _{x} \\
-\omega _{y} & \omega _{x} & 0
\end{array}
\right) .
\]

The \emph{exponential map}, $\exp :\mathfrak{se}(3)\rightarrow
SE(3)$, is given by
\[
\exp \left(
\begin{array}{cc}
\omega & v \\
0 & 0
\end{array}
\right) =\left(
\begin{array}{cc}
\exp (\omega) & Av \\
0 & 1
\end{array}
\right) ,
\]
where

\[
A=I+\frac{1-\cos \left\Vert \omega \right\Vert }{\left\Vert \omega
\right\Vert ^{2}}\omega+\frac{\left\Vert \omega \right\Vert -\sin
\left\Vert \omega \right\Vert }{\left\Vert \omega \right\Vert
^{3}} \omega^{2},
\]
and $\exp (\omega)$ is given by the \emph{Rodriguez' formula},
\[
\exp (\omega)=I+\frac{\sin \left\Vert \omega \right\Vert }{%
\left\Vert \omega \right\Vert }\omega+\frac{1-\cos \left\Vert
\omega \right\Vert }{\left\Vert \omega \right\Vert
^{2}}\omega^{2}.
\]

\end{document}